\documentclass[twocolumn,nofootinbib]{revtex4}

\usepackage{amsmath,latexsym,amssymb}
\newcommand{\be}{\begin{eqnarray}}
\newcommand{\ee}{\end{eqnarray}}
\newcommand{\bi}{\bibitem}

\newcommand{\mplq}{m_{Pl}^2}

\newcommand \bra {\langle}
\newcommand \ket {\rangle}

\begin{document}
\input epsf


\title{
\bf 
A Way to  Dynamically Overcome the Cosmological Constant Problem
}
\author{\bf Denis Comelli }
\address{\it INFN - Sezione di Ferrara, 
via Saragat 3,  Ferrara Italy  }

\date{\today}
\noindent
 
\begin{abstract}

The Cosmological Constant  problem can be  solved once we require
that  the full standard Einstein Hilbert lagrangian, 
 {\it gravity plus matter}, is multiplied by a  total derivative.
We analyze  such a picture 
writing the  total derivative  as the covariant gradient of a
new  vector  field ($b_{\mu}$).
The dynamics of this $b_{\mu}$ 
field can  play a  key role in the explanation of  the present cosmological
acceleration of the Universe.
\end{abstract}

\vskip1pc

\maketitle
\vspace{2.cm}

\section{\it \bf  Introduction.}

There are numerous suggestions in the literature for modification of the
classical Einstein Hilbert (EH) eqs of motion of general 
relativity 
($G_{\mu\nu}\equiv R_{\mu\nu}-
\frac{g_{\mu\nu}}{2}R=T_{\mu\nu}/\mplq $)
in connection with the solution of
the Cosmological Constant (CC) problem \cite{weinb},\cite{CC}.
Many approaches concentrate their attention on possible modifications
of the the world volume ($\int d^4x \sqrt{g}$),
 in particular, we will  briefly review the main
features of Unimodular Gravity and Two Measure Theory (TMT) whose
features are most similar to our findings.

In unimodular gravity, see \cite{uni}, the salient point is the non
dynamical character of the determinant of the metric tensor
 $g=det||g_{\alpha\beta}||$ that is frozen  to a constant value,
 leaving a theory  invariant only under volume
preserving general coordinate transformations \cite{unimodular}.
In absence of matter the  action can be written as
\be
S_{Un}=\int d^4x \sqrt{g}\,\,(\,R+\Lambda)+\lambda(g-1)\;
\ee
where $\lambda$ is a lagrange multiplier that fix the constraint $g=1$
 and $\Lambda$ the CC.
The unimodular eqs of motion are
\be\label{uni1}
R_{\mu\nu}-\frac{1}{4}g_{\mu\nu}R=0
\rightarrow 
\nabla_{\mu}R=0
\ee
where there is no trace of  the initial $\Lambda$ CC;
 however the solution remains  of  DeSitter type:
$ R_{\mu\nu}=\frac{1}{4}\;g_{\mu\nu}\;\Lambda'$
with $\Lambda'=const$ a new CC   different from the original one
($\Lambda$), coming from the boundary conditions of  eqs (\ref{uni1}) 
(the same conclusions are obtained in presence of  matter).
This, strictly speaking, does not solve the problem of the CC but 
 it changes the perspective and allows one to think of the CC as a non
 dynamical entity.

An other interesting approach is the 
 Two Measure Theory, introduced and fully developed by
Guendelman and Kaganovich \cite{gk}.  
In this case the action is of the form 
\be
S=\int d^4x\;\Phi \;{\cal L}_1+\int d^4x\;\sqrt{g} \;{\cal L}_2
\ee
with two lagrangians ${\cal L}_{1,2}$ functions of all matter fields, the
metric, the connection (the theory is defined  in first order
 Palatini formalism) and two different volume elements 
($\sqrt{g} \,d^4x$  and  $\Phi\, d^4x$)
where  $\Phi$ is a scalar density:
\be
\Phi=\epsilon^{\mu\nu\rho\sigma}\epsilon_{abcd}
\partial_{\mu}\phi_{a}
\partial_{\nu}\phi_{b}
\partial_{\rho}\phi_{c}
\partial_{\sigma}\phi_{d}
\ee
that results a total derivative
of the four fundamentals $\phi_{a}$ (a=1,2,3,4) scalar fields.
The outcome of the extra measure is the presence of a new scalar field
$\zeta=\frac{\Phi}{\sqrt{g}}$ which couple, after a conformal
transformation, in a non trivial way  to fermions and 
to an effective scalar  potential that is
 automatically minimized into a state with zero CC without tuning of
the parameters. 

To our knowledge only few attempts to modify directly the structure 
of the  world volume are present in literature, see for example
\cite{wil} and \cite{dc}, also if somehow 
they are not devoted to the CC problem.

\vspace{0.2cm}

Now let us come to our proposal.
The key ideas  are based on  basic properties:
one is related to  the covariant divergence of a vector field $b_{\mu}$:
$\sqrt{g}\;
\nabla^{\alpha}\;b_{\alpha}=\partial^{\alpha}(\sqrt{g}\;b_{\alpha})$
being a total derivative  and the other is the fact that
a shift in the total lagrangian ${\cal L}\rightarrow {\cal L}+ \;const $
has no effect on the gravitational eqs of motion
once we multiply everything by a total derivative.
The combination of this two simple ideas can be summarized in the following non 
dynamical  action:
\be
\int d^4x \sqrt{g} \;\nabla_{\alpha}b^{\alpha}\;{\Lambda}={\Lambda}
\int d^4x \;\partial_{\alpha} (\sqrt{g}\,\,b^{\alpha})=0
\ee

This motivate us  to \underline{\it postulate} 
that the total Lagrangian (including gravity and matter) 
$\bar{\cal L}$ of our world
has to be multiplied by a total derivative 
\be
S=\int d^4x \;\sqrt{g} \;\;\nabla^{\alpha}b_{\alpha}\;\;\bar{\cal L}
\ee
in order to be  automatically  insensitive to any CC  coming from
the  processes of renormalization or phase transitions.
For the rest of the paper we use 
$\chi\equiv\nabla^{\alpha}b_{\alpha}$, 
having in mind the fact that we can introduce  others total derivative terms as
 for example $\nabla^2 \phi$ with $\phi$ a scalar field.
The above postulate assumes implicitly that,
 if we start with a bare lagrangian $\bar{\cal  L}_0$,
 the process of renormalization, generating the renormalized
 lagrangian 
$\bar{\cal L}$, with the corresponding CC
 $\Lambda$, conserves the original structure:
\be
\int d^4x \sqrt{g} \; \;\chi\;
\bar{\cal  L}_0 \!\!\!\!\!\!\!\!\!\!
\underbrace{\Rightarrow}_{\rm Renormalization}
\!\!\!\!\!\!\!\!\!\!
\int d^4x \sqrt{g} \;\chi
\;(\bar{\cal L}+\Lambda)
\ee
This result can be obtained if some symmetry principle can be worked
 out.
We note, for example, that the full action is \underline{\it odd} under the 
discrete symmetry $x_{\mu}\, \rightarrow -\, x_{\mu}$ 
for which 
$\chi\rightarrow -\chi$ , and $R\rightarrow R $. The matter lagrangian
 is naturally even for fields with an even power of derivatives
(scalars and vectors) while, in order to accommodate also the
 fermionic 
fields which have only one derivative, we  can ask also a non trivial 
transformation for
the vierbien fields: $e^{\alpha}_{\mu}\, \rightarrow -\, e^{\alpha}_{\mu}$.

The above statements can be translated also in a  postulated new 
modified world volume:
\be
\int d^4x\;\sqrt{g}\rightarrow \int d^4x\;\sqrt{g}\;\chi
\ee 
We note also that 
$\int d^4x\,\sqrt{g}\,\chi\,R $ cannot be modified in an Einstein form through a 
 conformal transformation 
 contrary to the case in
which $\chi$ is a simple scalar field.

The subject of our analysis is the following action
\be\label{s}
S=\int d^4x \;\sqrt{g} \;\chi  \;(\mplq\;R-{\cal L})
\ee
where ${\cal L}\equiv {\cal L}_m+{\cal L}_b$
  contains the usual matter Lagrangian $ {\cal L}_m $ and the dynamics of the
$b_{\mu}$ field (${\cal L}_b$).

We stress  that many other scenarios can be implemented along
the same lines, for instance 
$\int d^4x \sqrt{g} \;(\chi_1\,R-\chi_2
\;  {\cal L})$
with the presence (or the absence) of different $\chi_i$.
We note that many formulas are comparables to the ones  in the TMT
\cite{gk} also if the main assumptions  result quite differents 
(for example our framework is a Riemannian geometry).

The eqs of motion from  action (\ref{s}) are:
\begin{itemize}
\item For the vector field :
\be\label{eqvec}
\nabla_{\mu}(\mplq\,R\!-\!{\cal L})=
\nabla^{\alpha}\!\left(\chi\frac{\partial {\cal
    L}_b}{\partial \nabla^{\alpha}b^{\mu}}\right)\!-\!\chi\frac{\partial {\cal
    L}_b}{\partial b^{\mu}} \equiv J_{\mu}^{(b)}
\ee

\item For the matter fields ($\phi$):
\be\label{eqmatt}
\nabla^{\alpha}\left(\frac{\partial {\cal
    L}_m}{\partial \nabla^{\alpha}\phi}\right)-\frac{\partial {\cal
    L}_m}{\partial\phi}=-\frac{\nabla^{\alpha}\chi}{\chi}
\frac{\partial 
{\cal L}_m}{\partial \nabla^{\alpha}\phi}
\ee

\item For gravity :
\be\label{eqgrav}\nonumber
\chi\,&&
\left( \mplq\,R_{\mu\nu}-\frac{\partial {\cal L}}{\partial g^{\mu\nu}}\right)=
  \frac{b_{\mu} J_{\nu}^{(b)}+ b_{\nu} J_{\mu}^{(b)}}{2}- \\
&&\frac{g_{\mu\nu}}{2}b^{\alpha} \, J_{\alpha}^{(b)}-
\mplq
\left(g_{\mu\nu}\nabla^2-\nabla_{\mu}\nabla_{\nu} \right)
\chi
\ee

\end{itemize}

where we used eq (\ref{eqvec}) to get rid of the
terms $\nabla_{\nu}\left(\mplq\,R-{\cal L} \right)$ as function of the
current $J^{(b)}_{\mu}$.  
Eq (\ref{eqgrav}) can be further simplified  after the definition of the tensor
\be
\tilde{T}_{\mu\nu}\equiv\frac{\partial{\cal L}}{\partial
  g^{\mu\nu}}
\ee
 in the more usual form
\be\nonumber 
\mplq &&\left( R_{\mu\nu}\!-\frac{g_{\mu\nu}}{2}R\right) =\tilde
T_{\mu\nu}-\frac{g_{\mu\nu}}{2} \tilde T+
\frac{1}{2\,\chi}
\left(b_{\mu}J_{\nu}^{(b)} +\right.\\ && \left. b_{\nu}J_{\mu}^{(b)}\right)+ 
\frac{\mplq }{2\,\chi} \left(g_{\mu\nu}\nabla^2\!+2\,\,\nabla_{\mu}\,\nabla_{\nu} \right)
\chi
\ee
that can also be  rewritten in a compact form as
\be\label{newgr}
G_{\mu\nu}= \frac{1}{\mplq } 
\left(T^{(m)eff}_{\mu\nu}+  T^{(b)eff}_{\mu\nu} \right)  \label{eqgg}
\ee
The  $ \tilde T^{}_{\mu\nu}$ tensor being related to  the usual 
energy momentum (EM) tensor by 
\be
T_{\mu\nu}\equiv \frac{\partial(\sqrt{g}{\cal L})}{\sqrt{g}\partial
  g^{\mu\nu}}=
-\frac{g_{\mu\nu}}{2}\;{\cal L}+\tilde{T}_{\mu\nu}
\ee  
allow us to write the new matter source of gravity  as 
\be
T^{(m)eff}_{\mu\nu}\!\equiv\!
\tilde T_{\mu\nu}^{(m)}\!-\!\frac{g_{\mu\nu}}{2} \tilde T^{(m)}\!=\!
T_{\mu\nu}^{(m)}\!-\!\frac{g_{\mu\nu}}{2} 
( T^{(m)}+{\cal L}_m) 
\ee
 while  the vector field turn on the gravitational field by means of the tensor
\be\nonumber
 T^{(b)eff}_{\mu\nu}\equiv
&& \tilde{T}_{\mu\nu}^{(b)}-\frac{g_{\mu\nu}}{2} \tilde T^{(b)}+
\frac{1}{2\,\chi}
(b_{\mu}J_{\nu}^{(b)}+ b_{\nu}J_{\mu}^{(b)} )+
\\&&
\frac{\mplq }{2\,\chi}\left(\nabla^2\;g_{\mu\nu}+2\;\nabla_{\mu}\nabla_{\nu}\right)
 \chi
\ee

Finally we give the covariant conservation law for the matter EM tensor: 
\be\label{Tm}
\nabla^{\alpha}T_{\mu\alpha}^{(m)}=-\tilde T_{\mu\alpha}^{(m)}\;
\frac{\nabla^{\alpha}\;\chi}{\chi}
\ee
or in other terms:
$
\nabla^{\alpha}\left( T_{\mu\alpha}^{(m)}\,\chi\right )=-\frac{ {\cal L}_m}{2}\nabla_{\mu}\chi
$.
\vspace{0.2cm}

In order to be as simple as possible 
we take a  vector lagrangian that
  depends separately on
the combinations
 $\chi=\nabla^{\alpha}b_{\alpha}$ and
$b^2=b^{\alpha}b_{\alpha}$ :
\be
{\cal L}_b=f(\chi)+U(b^2)
\ee
with a  vector current $ J^{(b)}_{\mu}=\nabla_{\mu}(xf')-2\,\chi\, U'
\,b_{\mu} $   and  an effective EM tensor
\be\label{eff}
 T^{(b)eff}_{\mu\nu}=
-\frac{g_{\mu\nu}}{2}\left(\chi\, f'
+U'b^2 \right)-
U'b_{\mu}b_{\nu}
+\\\nonumber
\frac{\mplq }{2\,\chi}\left(\nabla^2\;g_{\mu\nu}+2\;\nabla_{\mu}\nabla_{\nu}\right) \chi
\ee
that inside eqs (\ref{eqvec},\ref{eqmatt},\ref{eqgrav}) generate  the
full dynamics of the system.

\section{\it \bf  Gravitational Sources  }

The new gravitational eqs (\ref{newgr}) demand at this point some
 comments.
 The new sources of gravity, $T_{\mu\nu}^{(m)eff}+T_{\mu\nu}^{(b)eff}$, 
show {\it strong departures} from 
the structure of the classical EH sources
($T_{\mu\nu}^{(m)}+T_{\mu\nu}^{(b)}$).

Focusing only on  matter,  we note that 
 the EM tensor is conserved  once 
$\chi= const$ (see eq \ref{Tm})  and 
$T^{(m)eff}_{\mu\nu}$
reduces to $T^{(m)}_{\mu\nu}$ only when  $g_{\mu\nu}\,(T^{(m)}+{\cal L}_m)\ll T^{(m)}_{\mu\nu}$.
The exact cancellation of this extra peace is a stringent  condition on matter lagrangian \cite{gk}
\be\label{laggk}
g^{\alpha \beta}\frac{\partial {\cal L}_m}{\partial g^{\alpha \beta}}-{\cal L}_m=0
\ee
which means that ${\cal L}_m$ is a homogeneous function of $g^{\alpha \beta}$ of degree one.

Other important aspect is 
the  role played by 
different kinds of matter sources that we classify in two classes:
 the point particle ($pp$) 
 and  the coherent field ($cf$) configurations sources.

In general, both of them  contribute to the dynamics of the system 
but the key point that will be elaborated in the following is the fact that 
point particle dynamics,
 contrary to coherent field configurations,
 has  some freedom in his theoretical  structure
that allow us to obtain phenomenological viable matter sources.

\subsection{\it \bf  Matter Point Particle Sources }

The classical geodesic eqs 
 for   point particles extremize the functional
$S_{pp}=\int d \tau \,{\cal L}_{pp}=-m\int d \tau\;\frac{d s}{d \tau}$,
with $ds=\sqrt{g^{\alpha\beta}dx_{\alpha}x_{\beta}}$ the world element.
 The corresponding  energy momentum tensor results 
$T_{\mu\nu}^{(pp)}=-\frac{1}{2}\int d \tau\,{\cal L}_{pp}
\,v_{\mu}\,v_{\nu} $ (with $v_{\mu}\equiv \frac{dx_{\mu}}{d \tau}$).
 In this case the ``anomalous'' term  counts
$g_{\mu\nu}({\cal L}_{pp}+T^{(pp)})=-\frac{g_{\mu\nu}}{2}\int d \tau\,{\cal L}_{pp}$
 and is never subdominant.
The key point  to evade such a non physical implication
 is the realization that for free point particles  a full variety
 of actions are possible \cite{gk}.
The special property of the above $S_{pp}$ is  reparametrization
invariance but in general if, for example, we use a generalized action  
${\cal S}_{pp}=-m \int d\tau F(\frac{ds}{d \tau})$ 
 with $F(\frac{ds}{d \tau})$ a generic
functional of the variable $(\frac{ds}{d \tau})$ and $\tau$ 
an affine parameter, in the EH context we 
generate the correct geodetic equations  and the correct structure of
the EM tensor.
In our case instead, different choices of $F$
generate unequivalent theories.
In fact now $T^{(pp)}_{\mu\nu}=\frac{m}{2}\int d\tau \;F'\,v_{\mu}\,v_{\nu} $ 
and 
$g_{\mu\nu}({\cal L}_{pp}+T^{(pp)})=  g_{\mu\nu}\;m\int d\tau \;
(-F+\frac{ F'}{2} )$ .

If we choose
 $F=\left(\frac{ds}{d\tau}\right)^2=g^{\alpha\beta}\frac{d x_{\alpha}}{d \tau}\frac{d x_{\beta}}{d \tau} $ we can get rid of the anomalous
 term \cite{gk} and  we obtain 
  $T^{(pp)eff}_{\mu\nu}=T^{(pp)}_{\mu\nu}$ (note that now the $pp$ lagrangian is satisfying 
eq (\ref{laggk})).   
In this way  the
 $pp$  dynamics  coincides with the EH case.

\subsection{\it \bf Matter Coherent Field Sources }

In order to work out the specific features of coherent field configurations
 we take  a simple  scalar field lagrangian
 \be\nonumber
 {\cal L}_m= g^{\alpha\beta}\partial_{\alpha}\phi\partial_{\beta}\phi-V(\phi)
\ee

The effective EM tensor results:
\be
T_{\mu\nu}^{eff(m)}=\partial_{\mu}\phi\partial_{\nu}\phi-
\frac{g_{\mu\nu}}{2}(\partial \phi)^2=T_{\mu\nu}-
\frac{g_{\mu\nu}}{2}V(\phi)
\ee
and the effective  energy  and pressure  density in a FRW universe 
with a scale factor 
$a\equiv a(t)$ result
\be\label{rm}
\rho_m^{eff}= \frac{\dot\phi^2}{2}+
\bra \left(\frac{\vec \nabla \phi}{\sqrt{2}a}\right)^2 \!\!
\ket ,
\;\;p_m^{eff}=\frac{\dot\phi^2}{2}-
\bra \left(\frac{\vec \nabla \phi}{\sqrt{6}a}\right)^2 \!\!
\ket 
\ee
where an average $<...>$ over spatial gradients is imposed in order to
 preserve the spatial homogeneity.

The relationship between the usual eq of state $w=\frac{p}{\rho}$ and
$w_{eff}=\frac{p_{eff}}{\rho_{eff}}$  during the various cosmological epochs is here synthesized
\be
\underbrace{(-1,\,-\frac{1}{3},\,0,\,\frac{1}{3},\,1)}_{w}\leftrightarrow
\underbrace{(Indef.\, ,\,-\frac{1}{3},\,1,\,\frac{1}{3},\,1)}_{w_{eff}}
\ee
where we note as usual Inflationary phase ($w=-1$) cannot be generated with the new dynamics and 
 the usual matter phase ($w=0$ with $\vec \nabla \phi=0$) of a classical
 oscillating coherent scalar field
is now replaced by a faster kinetic phase.
To evade such a conclusions it is in needed of a 
non trivial $g_{\alpha\beta}$ dependence in the interactions as it is
 the case for vector fields, like in the  $b_{\mu}$ dynamics.

It is then clear that all the
cosmological considerations related to the presence of dominant coherent fields
( like during the classical Inflationary period) have to be deeply reanalyzed.

\section{Vector Dynamics and CC }

Now let us consider the  vector field dynamics.
What we discover
is the fact that without  interactions (in particular $U= 0$)
 the system develops a CC completely unconstrained induced
by the boundary conditions in a way  similar  to  unimodular gravity.
 An appropriate choice of self interactions ($U\neq 0$)
instead  can dynamically constrain the system to calibrate his
energy densities.

The key point is  the vector  eq  (\ref{eqvec})
(see also eq (\ref{nabla})).

\vspace{0.2cm}

In the  case with \underline{ $U=0$} (or in
general when $J^{(b)}_{\mu}$ is zero)
eq (\ref{nabla}) results a total derivative
 and consequently a  new 
CC $\hat\Lambda $ term comes from the boundary conditions 
and fixes the following sum: 
$T^{(m)} +{\cal L}_m+\chi f'-f-3 \frac{\nabla^2\chi}{\chi}=
\hat\Lambda$
that backreacts on gravity as
\be\label{U0}
G_{\mu\nu}=\frac{
T^{(m)}_{\mu\nu}-\frac{g_{\mu\nu}}{2}(\hat\Lambda +f)}{\mplq}+
\frac{(\nabla_{\mu} \nabla_{\nu}-g_{\mu\nu}\nabla^2)\chi}{\chi}
\ee

 If we take the  derivative  interaction
 $f=\lambda(\chi-\bar\chi)$, with $\lambda$ a
lagrange multiplier and $\bar\chi$ a fixed background
we have the constraint $\chi=\bar\chi$  and we reduce
 eqs (\ref{U0})
to  the  EH ones  in presence of the CC $\hat \Lambda$.
The parallelism with unimodular  gravity  shows interesting analogies.

In  absence of interactions ( $U=f=0$) and of
 matter we find  a De Sitter Space with
$\mplq\,R=\hat\Lambda\rightarrow 
\nabla^2\,\chi+\frac{\hat \Lambda}{3}\chi=0$ so that 
$\chi(t)=\chi(0)\;e^{-H\,t}$
with $ H^2 =-\frac{\hat\Lambda }{12\,\mplq}$.

\vspace{0.2cm}

The opposite case with  \underline{$U\neq 0$}  ( or in
general when $J^{(b)}_{\mu}$ is non zero ) instead results much more
interesting.
In order to show explicitly the dynamical properties of the system
let us take a
FRW universe with a background vector field $b_{\mu}=(b(t),0,0,0)$.

The eqs of motion now result
\be
\label{FRW}\nonumber
U'b^2\!\!&=&\!\!\mplq\,H\!\left(2 \, H'+\frac{H\,\chi''}{\chi}
+(H'-  H)\frac{\chi'}{\chi}\right)+\!\rho_m+\!p_m\\  \nonumber
\chi f'&=&-6\,\mplq\,H\,\left(
 H' +H-H\frac{\chi'}{\chi}
\right)-\left({\cal L}_m+2\rho_m
\right)\\ 
 \rho_m' &+& 3(p_m+\rho_m)=-\frac{(2 \rho_m+{\cal L}_m)}{2}\frac{\chi'}{\chi}
\ee
with $\chi= \,H\,(b'+3b)$ and
where the apex $'$ are $x$ derivatives ( $x\equiv log \,a(t)$ ) .

 The choice  $f=\lambda\;\left(\chi-\bar \chi\right)$
allows us to obtain  informations in the asymptotic regime
($x\rightarrow \infty$) where we can safety neglect matter contributions.
The  system of eqs reduces to : $H'H=\frac{1}{2 \mplq}U'b^2$ plus
$H(b'+3\,b)=\bar\chi$  with the  DeSitter solutions:
\be
\left.U'( b^2)\; b^2\right|_{b=\bar b}=0, \;\; \;\;\bar H=\frac{\bar \chi}{3 \bar b}
\ee
so, the value of $\bar b$ minimizes the potential $U$ and 
the corresponding
Hubble time results  proportional to the ratio  $\bar
\chi/\bar b$.
As example taking $U=\alpha(b^2-m_b^2)^2$ we have $\bar b=m_b$. 
If we  fit this scenario with
the beginning of our present
cosmological acceleration  we obtain:
\be
\frac{\bar \chi}{ \bar b}\;\sim\;\sqrt{\frac{\rho_0}{\mplq}}\;\sim\; 10^{-41}\;GeV
\ee
that looks a  sort of fine tuning necessary to be consistent with the
 cosmological parameters describing our world.

The case with the potential $U=M^4\,log\, b^2$ is quite
 interesting because  the
combination $U'\,b^2$ is a constant.
With  no matter and  $f=0$, we obtain the exact solution 
\be
H^2= H_i^2+\frac{M^4}{2\,m_{pl}^2}(x-x_i)
\ee
with   $H(x_i)=H_i$.
Note that in this case the asymptotic value ($x\rightarrow \infty$) of $H$ is 
proportional to $M$, the scale of the potential $U$  and not to  a generic
initial constant as in the case with $U=0$.

In order to extend our analysis to physical scenario we have to
integrate numerically the system of differential eqs (\ref{FRW}).
We take $f=\lambda(\chi-\bar \chi)$  to reduce as much as
possible  the differential eq order
and we take two different $U$ potentials:
the first case with
 $U=\alpha\,( b^2-m_b^2)^2$ and then  $U=\alpha\, b^4$ choosing the
appropriate parameters that give, in both cases, the same matter content at
$x=0$.
We didn't care about possible multi degeneracy in the parameter space,
 but we take this examples as possible workable toy models.

For the matter content we take only the point particle energy momentum tensor
 that gives the usual  eq of state ($w=w_{eff}$)
 due to the modified point particle lagrangian.
Initial conditions are given in the deep radiation era.

Fixing  today the matter density $\Omega_m=0.3$ and the $b$-vector density $\Omega_{Vect}=0.7 $,  in the
first case we find  $\alpha=-2.8 \,10^{-10},\, \bar \chi=5.5\,10^{-51}\,GeV^2$ 
and $m_b=10^{-10}\,GeV$ while in the second case we find
 $\alpha= \,10^{-22},\, \bar \chi=8\,10^{-47}\,GeV^2$.
Following \cite{R}
we  compute also  the CMB shift parameter 
${\cal R}
=1.716\pm 0.062$
from WMAP  and the
 parameter ${\cal A}=0.469\pm 0.017$ associated with the determination of the
baryon acoustic peak, just to have a hint of the cosmological scenario 
that can be obtained in these models. 

\begin{figure}[htb]
\epsfxsize=3.3 in
\epsfysize=2.5 in
\begin{center}
\leavevmode
 \epsfbox{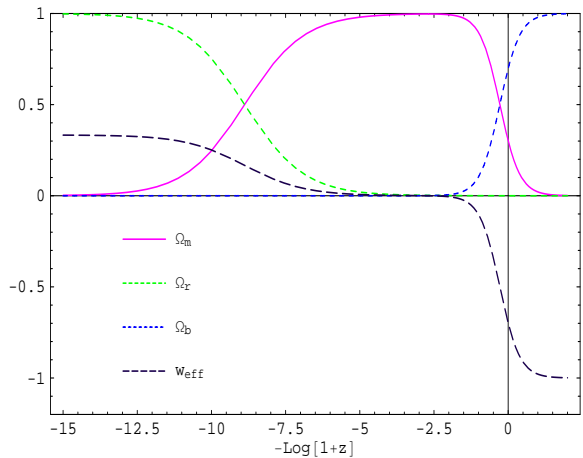}
\end{center}
\caption{
Plot of $\Omega_{m},\;\Omega_{r}$ (radiation density),$\; \Omega_{Vect} $
 and the effective eq of state $ w_{eff} $
as a function of the redshift $ x=-Log[1+z] $ for 
$ f=\lambda(\chi-\bar \chi)$, and $ U=\alpha\,(b^2-m_b^2)^2 $
with $\alpha=-2.8\, 10^{-10},\, \bar \chi=5.5\,10^{-51}\,GeV^2 $ and
$m_b=10^{-10}\,GeV $.
}
\end{figure}

\begin{figure}[htb]
\epsfxsize=3.3 in
\epsfysize=2.9 in
\begin{center}
\leavevmode
 \epsfbox{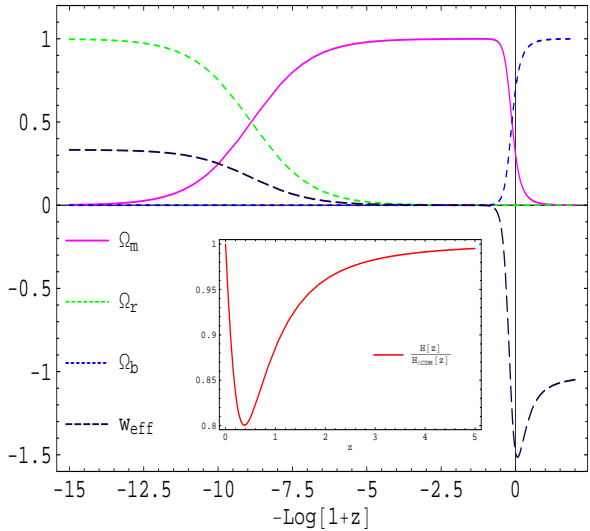 
}
\end{center}
\caption{
Plot of $ \Omega_{m},\;\Omega_{r}, \;\Omega_{Vect} $ and $ w_{eff} $
as a function of the redshift $ x=-Log[1+z] $ for 
$ f=\lambda(\chi-\bar \chi) $, and $ U=\alpha\,b^4 $
with $ \alpha= \,10^{-22},\, \bar \chi=8\,10^{-47}\,GeV^2$.
In the internal square we plot  also the ratio between the 
$z-$dependent Hubble parameter
of our model over the  $\Lambda$CDM scenario with the same 
late time $z=0$ values of $\Omega_m$ and $\Omega_{\Lambda}=\Omega_{Vect}$.
}
\end{figure}

For the potential $U=\alpha \,(b^2-m_b)^2$ 
we find  at $x=0$ an  effective eq of state $w_{eff}=-0.7$ with  
 ${\cal A}=0.49$ and ${\cal R}=1.75$ and a
 time dependent  Hubble expansion  identical to the normal
  $\Lambda$CDM 
scenario having the same 
late time ($x=0$) values of $\Omega_m$ and $\Omega_{\Lambda}=\Omega_{Vect}$:
$H_{\Lambda CDM}[z]=\sqrt{\Omega_m(1+z)^3+\Omega_{\Lambda} }$ (see fig.
1).

For 
$U=\alpha \,b^4$ 
we find  at $x=0$ an  effective eq of state $w_{eff}=-1.48$ with 
 ${\cal A}=0.58$ and ${\cal R}=1.85$
with a pronounced deviation from the  $\Lambda$CDM 
scenario in the present time (see fig 2),
 note also  the strange dependence of the effective eq of state 
that jump to $\sim -1.5$ around $z\sim 0$ and then recalibrate
 to $\sim -1$ in late time period.

A detailed analysis of of cosmological pictures obtained with these models 
(attractor points, etc. )has
to be work out.

As a final comment we note  that
the presented scenario doesn't fit with the main hypothesis of the 
{\it CC no-go Weinberg theorem} \cite{weinb} due 
to the fact that it requires 
{\it all fields  to be constant on the vacuum}.
We stress, in fact, that in order to have a non trivial dynamics 
we need $\chi=\nabla^{\alpha}b_{\alpha}\neq 0$  all times and this
simple fact lend wings to the mechanism of CC cancellation here described.
A similar feature is  present also in the TMT \cite{gk}.

Many open problems are still to be investigated, first of all
the presence of ``anomalous'' contributions in the sources of the EH eqs
that can generate phenomenological interesting dynamical deviations from EH.
It follows  the study of the inflationary mechanism and
 the  dynamics of the  linear perturbations than can deserve  surprises 
due to the higher derivative structure of the theory.

\vspace{0.2cm}

{\bf Acknowledgments}:
I would like to thanks  A. Dolgov   for stimulating discussions.

\vspace{0.5cm}

{\bf \large Appendix:}

\vspace{0.2cm}



Some notes about   vector dynamics:
\be \nonumber
\tilde T^{(b)}_{\mu\nu}=\frac{\partial{\cal L}_b}{\partial g^{\mu\nu}}=
\frac{1}{\chi}\left(
-\hat B_{\mu\nu}(\chi f')+\frac{g_{\mu\nu}}{2}
\nabla^{\alpha}(b_{\alpha}\chi f')\right)
+U'\,b_{\mu}b_{\nu}
\ee
where
$\hat B_{\mu\nu}(F)\equiv
\frac{1}{2}(b_{\mu}\nabla_{\nu}F+b_{\nu}\nabla_{\mu}F),\;
f'\equiv \frac{\partial f}{\partial \chi}$,

$\;U'\equiv \frac{\partial U}{\partial b^2}$.
   We give also the trace of eq (\ref{eqgrav})
\be\label{trace}\nonumber
\mplq \,R=T^{(m)}+2{\cal L}_m+2\,\chi\, f'
+3\, U'b^2 -3\,\mplq\,  \frac{\nabla^2 \chi}{\chi}
\ee
that with the vectorial eqs of motion gives    
\be\label{nabla}
&\nabla_{\mu}&\left( T^{(m)}\!\!+{\cal L}_m+ \nonumber
\chi\, f'-f+
3 \,U'\,b^2-U-\right.\\&&\left.
3\,\mplq \frac{ \nabla^2 \chi}{\chi}\right)=
-2 \,\chi \,U' \,b_{\mu}
\ee

\end{document}